\documentclass[reprint,amsfonts, amssymb, amsmath, showkeys, nofootinbib]{revtex4-2}

\usepackage{graphicx}%
\usepackage{dcolumn}%
\usepackage{bm}%
\usepackage[english]{babel}
\usepackage[utf8]{inputenc}
\usepackage{subfigure}
\usepackage{dirtytalk}
\usepackage[colorinlistoftodos, color=green!40, prependcaption]{todonotes}
\usepackage{amsthm}
\usepackage{mathtools}
\usepackage{physics}
\usepackage{xcolor}
\usepackage{adjustbox}
\usepackage{placeins}
\usepackage[T1]{fontenc}
\usepackage{lipsum}
\usepackage{csquotes}
\usepackage[pdftex, pdftitle={Article}, pdfauthor={Author}]{hyperref} %
\def\mpchinv{$ \ h^{-1}\mathrm{Mpc}$~}

\newcommand{\utkarsh}[1]{\textcolor{purple}{(Utkarsh: #1)}}

\newcommand{\comment}[1]{}
\newcommand{\ie}{\textit{i}.\textit{e}.}

\def\x{{\bf x}}
\def\k{{\bf k}}

\def\fnl{f_{NL}}

\def\nbody{$N$-body }

\begin{document}
\title{Robust Neural Network-Enhanced Estimation of Local Primordial Non-Gaussianity}

\author{Utkarsh Giri}
    \email[Correspondence email address: ]{ugiri@wisc.edu}%
    \affiliation{University of Wisconsin-Madison, Madison, Wisconsin, USA}
    
\author{Moritz M\"unchmeyer}
    \affiliation{University of Wisconsin-Madison, Madison, Wisconsin, USA}

\author{Kendrick M. Smith}
    \affiliation{Perimeter Institute for Theoretical Physics, Waterloo, Ontario, CA}

\date{\today} %

\begin{abstract}

When applied to the non-linear matter distribution of the universe, neural networks have been shown to be very statistically sensitive probes of cosmological parameters, such as the linear perturbation amplitude $\sigma_8$. 
However, when used as a ``black box'', neural networks are not robust to baryonic uncertainty.
We propose a robust architecture for constraining primordial non-Gaussianity $\fnl$, by training a neural network to locally estimate $\sigma_8$, and correlating these local estimates with the large-scale density field.
We apply our method to $N$-body simulations, and show that $\sigma(\fnl)$ is 3.5 times better than the constraint obtained from a standard halo-based approach.
We show that our method has the same robustness property as large-scale halo bias: baryonic physics can change the normalization of the estimated $\fnl$, but cannot change whether $\fnl$ is detected.

\end{abstract}

\keywords{non-Gaussianity, neural networks, large-scale structure}

\maketitle

\section{Introduction} 
\label{sec:introduction}
 
Observations of cosmological perturbations on quasilinear scales, with the cosmic microwave background (CMB) and large scale structure (LSS), have resulted in precise measurements of fundamental cosmological parameters. 
However, quasilinear scales contain only a small fraction of the theoretically accessible information.
On smaller scales, traditional $N$-point correlation function analysis becomes sub-optimal, and inference methods such as forward modelling \cite{Seljak:2017rmr,2013MNRAS.432..894J} or machine learning \cite{Ravanbakhsh:2017bbi,Villaescusa-Navarro:2021pkb,Lazanu:2021tdl,Villaescusa-Navarro:2021cni,Hortua:2021vvj} can provide stronger constraints. 

Here, we will focus on inference of the important primordial physics parameter $\fnl$, which arises in multi-field models of inflation \cite{Linde_1997, Dvali_2004, Bartolo_2004, Biagetti_2019}.
In such models, the primordial Bardeen potential $\Phi(\x)$ can be parameterized as:
\begin{equation}
\Phi(\x) = \Phi_G(\x) + \fnl(\Phi_G(\x)^2 - \langle \Phi_G^2 \rangle)  \label{eq:pbs_fnl}
\end{equation}
where $\Phi_G$ is a Gaussian field and $\fnl$ quantifies the level of non-Gaussianity. Constraining $\fnl<1$ is a major goal of upcoming galaxy surveys \cite{Alvarez:2014vva}.

Several statistical approaches for estimating $\fnl$ in LSS have been proposed, including the squeezed bispectrum \cite{MoradinezhadDizgah:2020whw} and the scale-dependent power spectrum approach \cite{Dalal:2007cu, Slosar:2008hx} together with the idea of sample variance cancellation using a variety of probes \cite{Seljak:2008xr, Smith:2018bpn, Munchmeyer:2018eey, Giri:2020pkk}.

In this paper, we will propose a neural network (NN) based approach to estimating $\fnl$.
We will show that our NN-based analysis obtains significantly smaller error bars than an analysis based on large-scale matter and halo fields.
This is perhaps unsuprising since a neural network with enough capacity, trained on enough simulations, should give {\em statistically} optimal parameter constraints.

However, {\em robustness} is a central challenge for neural networks. 
As is widely appreciated, simulations on small scales suffer from baryonic feedback uncertainties and are not reliable at the sub-percent level accuracy required to tighten parameter bounds. 
For example, an NN trained to measure $\sigma_8$ on one set of simulations will likely give incorrect results on different simulations or on real data \cite{Villanueva-Domingo:2022rvn,Villaescusa-Navarro:2020rxg}.

This problem is not unique to NN-based methods.
In particular, the average dark matter halo density $\bar{n}_h$ is statistically a precise probe of $\sigma_8$, but it is not robust because $\bar n_h$ is sensitive to uncertain local physics. 
However, {\em anisotropy} in the halo field can be used to place robust constraints on $\fnl$.
A famous result \cite{Dalal:2007cu} states that if $\fnl\ne 0$, the halo bias $b_h(k)$ contains a characteristic $1/k^2$ term, which cannot be induced by local physics.

The idea of this paper is to replace the halo field by an NN-based local estimate $\pi(\x)$ of $\sigma_8$.
As in the halo case, $\bar\pi$ is not robust as an absolute measurement of $\sigma_8$, but the bias $b_\pi(k)$ contains a term proportional to $\fnl/k^2$, which can be used to place robust constraints on $\fnl$.
We will show that this approach combines the statistical power of neural networks with the robustness of the traditional halo-based analysis.

\section{Formalism} 
\label{sec:formalism}

In an $\fnl$ cosmology, the halo bias $b_h(k)$ contains a term proportional to $\fnl/k^2$ \cite{Dalal:2007cu}.
In the next few paragraphs, we review the derivation, in a language which will generalize to NN-based observables.

If $\fnl\ne 0$, the locally observed amplitude of short-wavelength modes is a function $\sigma_8^{\rm loc}(\x)$ of position.
On large scales $\k_L \rightarrow 0$, anisotropy in $\sigma_8^{\rm loc}$ is related to the potential $\Phi$ by:
\begin{equation}
\frac{\sigma_8^{\rm loc}(\k_L)}{\bar\sigma_8} = 2 \fnl  \, \Phi(\k_L)
\label{eq:pbs_sigma8_loc}
\end{equation}
The coupling between large and small scales described by Eq.\ (\ref{eq:pbs_sigma8_loc}) arises from the term $f_{NL} \Phi_G^2$ in Eq.\ (\ref{eq:pbs_fnl}), which mixes scales.
For a formal derivation of (\ref{eq:pbs_sigma8_loc}), see \cite{Slosar_2008, Biagetti_2019}.

On large scales, the local halo abundance $\delta_h$ is sensitive to both $\sigma_8^{\rm loc}$ and the matter overdensity $\delta_m$:
\begin{equation}
\delta_h(\x) = b^G_h \delta_m(\x) 
  + \frac{1}{2} b^{NG}_h \log\left( \frac{\sigma_8^{\rm loc}(\x)}{\bar\sigma_8} \right) 
  + (\mbox{noise})
\end{equation}
The first term $b^G_h \delta_m$ is the usual (Gaussian) halo bias, and the second term $b^{NG}_h \log(\sigma_8^{\rm loc})$ is a non-Gaussian bias term.
The coefficient $b^{NG}_h$ is the derivative of the halo density $\bar n_h$ with respect to the cosmological parameter $\sigma_8$ \cite{Slosar_2008,Baldauf_2011,Desjacques:2016bnm,Biagetti_2017}:
\begin{equation}
b^{NG}_h = 2 \frac{\partial \log \bar n_h}{\partial\log\sigma_8}
\label{eq:bh_ng}
\end{equation}
(Sometimes the approximation $b^{NG}_h \approx 2 \delta_c (b^G_h-1)$ is used, but we will not use it in this paper.)

On linear scales, the density field $\delta_m$ and potential $\Phi$ are related via the Fourier-space Poisson equation $\delta_m(\k) = \alpha(\k,z)\Phi(\k)$ \cite{Dodelson:2003ft}. Here, $\alpha(k,z)$ is given by %
\begin{equation}
\alpha(k,z) \equiv \frac{2k^2 T(k) D(z)}{3\Omega_m H_0^2}
\label{eq:alpha_def}
\end{equation}
Combining Eqs.\ (\ref{eq:pbs_sigma8_loc})--(\ref{eq:alpha_def}), we derive the NG halo bias:
\begin{equation}
\delta_h(\k_L) = b_h(k_L) \delta_m(\k_L) + (\mbox{Poisson noise})
\end{equation}
where:
\begin{equation}
b_h(k) = b^G_h + b^{NG}_h \frac{\fnl}{\alpha(k,z)}
\end{equation}
In this paper, we generalize the preceding results as follows.
First we note that they do not depend on any specific properties of halos, other than the local halo abundance $n_h(\x)$ being sensitive to $\sigma_8^{\rm loc}(\x)$.
Any field $\pi(\x)$ which is derived from the nonlinear density field, in a reasonably local way, should have the same property.
We propose constructing a field $\pi(\x)$ using a neural network trained to maximize sensitivity to $\sigma_8^{\rm loc}$.

We assume that our input data consists of the nonlinear density field $\delta_m(\x)$ as a 3-d pixelized map at $\sim$2 Mpc resolution without noise.
Thus, our constraints on $\fnl$ should be interpreted as information content in principle, given complete information at a specified resolution.
In future work, we plan to apply our approach to simulated galaxy catalogs, which are more representative of real data.

\begin{figure*}
\includegraphics[width=0.80\linewidth]{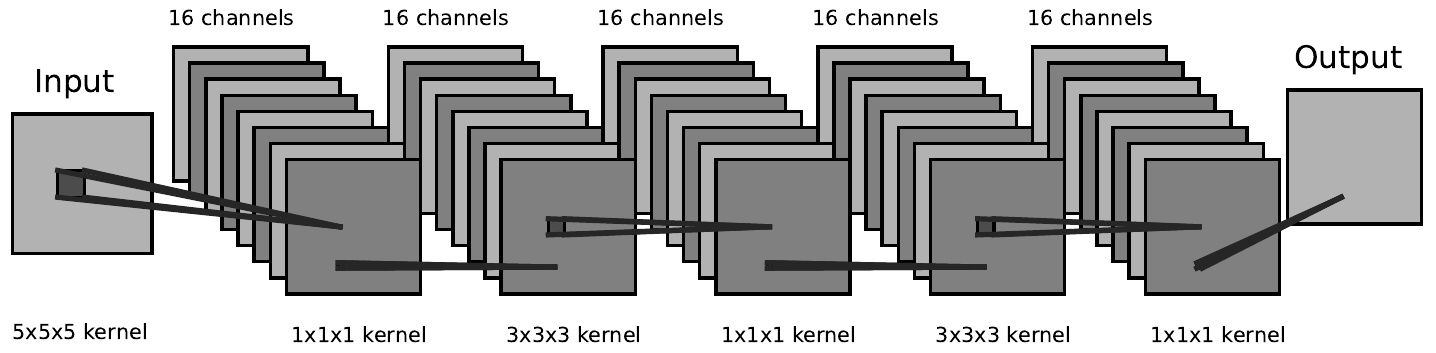}
\caption{A 2D schematic representation of our CNN architecture.
The input is the nonlinear 3-d density field $\delta_m(\x)$ from an $N$-body simulation, and we train the network so that the output 3-d field $\pi(\x)$ is an estimate of $\sigma_8$.
The total receptive field size is $(9 \times 9 \times 9)$ voxels, equivalent to $(18\ h^{-1} \mbox{Mpc})^3$.
Each convolution except the last is followed by a ReLU activation function. 
Each grey square represents a logical array of size $(512\times 512 \times 512)$ with periodic boundary conditions.
However, as an implementation detail to reduce GPU memory usage, we divide the simulation volume into slightly overlapping subvolumes which can be processed independently.
}
\label{fig:schematic_cnn}
\end{figure*}

We define a field $\pi(\x)$ by applying a CNN to $\delta_m(\x)$ (Figure \ref{fig:schematic_cnn}).
The CNN has a small receptive field ($\sim$18 $h^{-1}\, $Mpc), so that the output field $\pi(\x)$ is fairly local in the input field $\delta_m(\x)$.
As described in \S\ref{sec:pipeline}, we train the CNN so that $\pi(\x)$ is an estimate of $\sigma_8^{\rm loc}(\x)$ with low statistical noise.

Following the logic above for halos, we make the following predictions for the behavior of $\pi(\x)$ on large scales.
First, we predict that the matter-$\pi$ and $\pi$-$\pi$ power spectra are given by:
\begin{align}
P_{m\pi}(k) &= b_\pi(k) P_{mm}(k) \label{eq:P_m_pi} \\
P_{\pi\pi}(k) &= b_\pi(k)^2 P_{mm}(k) + N_{\pi\pi}  \label{eq:P_pi_pi}
\end{align}
where the linear bias $b_\pi(k)$ is the sum of Gaussian (constant) and non-Gaussian terms:
\begin{equation}
b_\pi(k) = b^G_\pi + b^{NG}_\pi \frac{f_{NL}}{\alpha(k,z)}
\label{eq:bpi_general}
\end{equation}
We also predict that the non-Gaussian bias $b^{NG}_\pi$ is related to the $\sigma_8$ dependence of the mean $\pi$-field:
\begin{equation}
b^{NG}_\pi = 2 \frac{\partial\bar\pi}{\partial\log\sigma_8}
\label{eq:bngpi_general}
\end{equation}
Finally, we predict that the noise $N_{\pi\pi}$ defined in (\ref{eq:P_pi_pi}) is constant in $k$.
In \S\ref{sec:mcmc}, we will verify these predictions and show that they lead to strong constraints on $\fnl$.

\section{Neural Network} 
\label{sec:pipeline}

\vskip5pt
\noindent
{\bf  Architecture:} 
Our neural network uses a fully convolutional, sliding-window architecture with a total of 16433 parameters (Figure \ref{fig:schematic_cnn}).
The network takes the 3D matter density field $\delta_m(\x)$ from an $N$-body simulation, and produces an output field $\pi(\x)$ with the same resolution as the input. 
We use small convolution kernels, including several layers with $(1\times 1\times 1)$ kernels, so that the total receptive field of the network will be small (18 $h^{-1}\,$Mpc).
The size of the receptive field limits the scales which the neural network can use for estimating $\pi$ and thus enforces locality.
We leave systematic exploration of neural network architecture to future work.

\vskip5pt
\noindent
{\bf  Simulations:} 
We want to train the network so that its output field $\pi(\x)$ is an optimal estimate of $\sigma_8$.
To do this, we need a training set of \nbody simulations with multiple values of $\sigma_8$.
We use the \verb|s8_p| and \verb|s8_m| datasets from the \texttt{Quijote} simulations \cite{Villaescusa-Navarro:2019bje}, with $\sigma_8=0.849$ and 0.819 respectively.
The remaining cosmological parameters are $\Omega_{\rm m}=0.3175$, $\Omega_{\rm b}=0.049$, $h=0.6711$, $n_s=0.9624$, and $w=-1$. 
Each dataset contains 400 collisionless simulations.
Each simulation has $512^3$ particles and volume $(1\ h^{-1} \mbox{Gpc})^3$. For each simulation, we inpaint particles from the $z=0$ snapshot on a $512^3$ 3D mesh using the Cloud-in-Cell algorithm implemented in \texttt{nbodykit}. This produces a voxelized 3D matter density field $\delta_m(\x)$, which we save to disk for neural network training. We require only two values of $\sigma_8$ in the training data rather than a continuum, because the variance of $\pi(\x)$ per receptive field is much larger than the difference between the two $\sigma_8$ values.

\vskip5pt
\noindent
{\bf Loss function and optimizer:}
For each simulation, let $\pi(\x)$ be the CNN output, and let $\sigma_8^{\rm true}$ be the value of $\sigma_8$ in the simulation. We define the loss function:
\begin{equation}
\mathcal{L} = \Bigg[
\Bigg(
\frac{1}{N_{\rm voxels}} \sum_{{\rm voxels}\ \x} \pi(\x) 
\Bigg) - \sigma_8^{\rm true}
\Bigg]^2
\label{eq:loss}
\end{equation}
Intuitively, minimizing $\mathcal{L}$ should produce an output field $\pi(\x)$ which is an optimal estimate of $\sigma_8$, by minimizing the difference between $\sigma_8^{\rm true}$ and the spatially averaged $\pi$-field. 

We use the Adam optimizer \cite{2014arXiv1412.6980K} with a learning rate of $(5 \times 10^{-5})$ to minimize the loss function. The learning rate is reduced by a factor of 0.7 whenever the loss fails to register any improvement for 5 successive epochs of training. The architecture is implemented in \texttt{PyTorch} \cite{https://doi.org/10.48550/arxiv.1912.01703} and uses \texttt{PyTorch-lightning} \cite{falcon2019pytorch} for high level interfacing with mixed precision training \cite{https://doi.org/10.48550/arxiv.1710.03740}.

We find that the overall normalization $W$ and additive bias $b$ of the NN are slow to converge, so as a final training step, we fix all parameters except $(W,b)$, and minimize the loss (\ref{eq:loss}).
This minimization can be done exactly in a single epoch, since the loss is a quadratic function of $(W,b)$.

\begin{figure}
    \centering
    \includegraphics[width=0.48\textwidth]{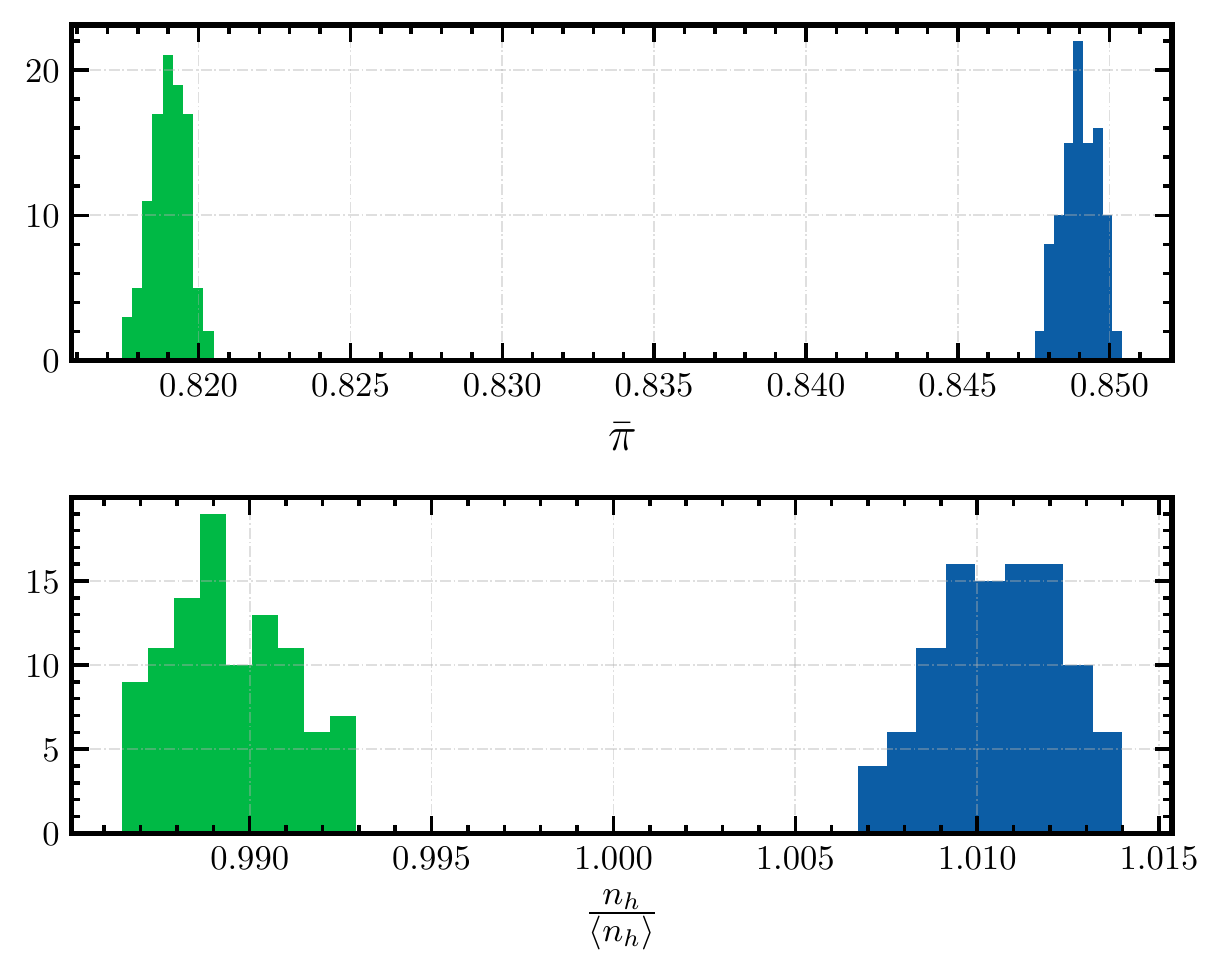}
    \caption{Estimating $\sigma_8$ using the neural network from Figure\ \ref{fig:schematic_cnn}.
    {\em Top panel.} Histogrammed NN estimates $\bar\pi = V_{\rm box}^{-1} \int_{\x} \pi(\x)$ on a test set of simulations with $\sigma_8=0.819$ (green) and $\sigma_8=0.849$ (blue).
    {\em Bottom panel.} Histogrammed halo counts from the same test set, showing worse statistical separation between $\sigma_8$ values than the NN.}
    \label{fig:sigma8_histograms}
\end{figure}

\comment{
\begin{figure}
    \centering
    \includegraphics[width=0.48\textwidth]{plots/projection_plot_10_percent.eps}
    \caption{2D slice of the $\ln(\delta_m)$ and $\ln(\pi)$ fields projected over 50\mpchinv along the z-axis. Large-scale correlations between the fields is visually evident.}
    \label{fig:my_label}
\end{figure}
}

\vskip5pt
\noindent
{\bf Validation:}
Our NN has been trained so that the spatially averaged $\pi$ field $\bar\pi = V_{\rm box}^{-1} \int_{\x} \pi(\x)$ is an estimate of $\sigma_8$ with lowest possible noise.
In the top panel of Figure \ref{fig:sigma8_histograms}, we verify this statement, by evaluating $\bar\pi$
on a test set of 100+100 simulations with $\sigma_8 \in \{ 0.819, 0.849 \}$.
We see that the network recovers the correct value of $\sigma_8$, and that the NN obtains better statistical separation between $\sigma_8$ values than counting halos (bottom panel).

\begin{figure}
    \centering
    \includegraphics[width=0.48\textwidth]{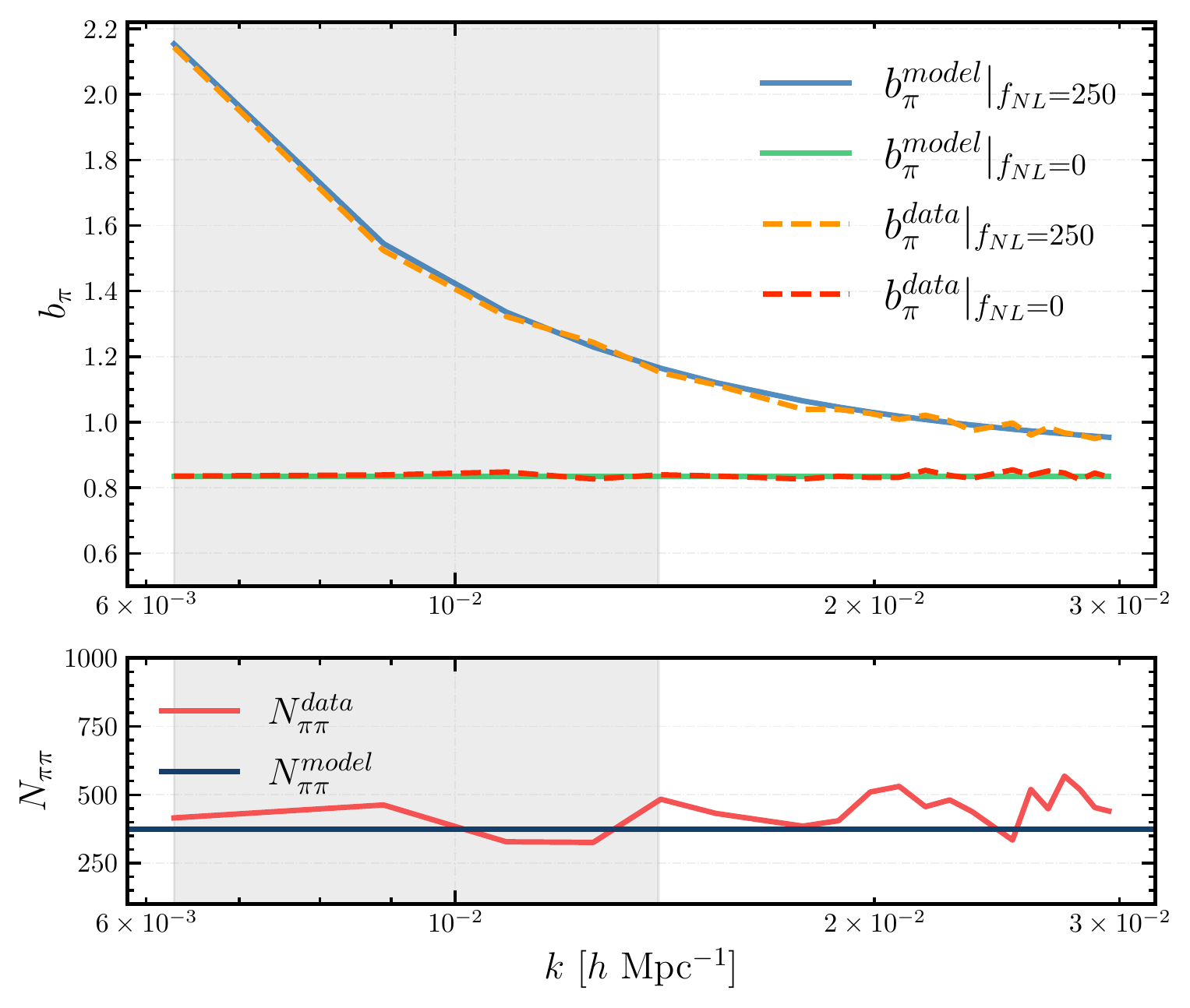}
    \caption{\emph{Top panel.} Bias model (\ref{eq:bpi_specific}) for the neural network output field $\pi(\x)$, compared to the empirical bias $P_{m\pi}(k)/P_{mm}(k)$ from simulation, for $\fnl \in \{0,250\}$. \emph{Bottom panel.} Power spectrum of the residual field $\epsilon(\k) = \pi(\k) - b_\pi(k) \delta_m(\k)$ compared to a best-fit constant $N_\pi$.
    Throughout this figure, best-fit model parameters $(b^G_\pi, N_\pi)$ are obtained from the MCMC pipeline described in \S\ref{sec:mcmc}, with $k_{\rm max}=0.014$ $h\, \mathrm{Mpc}^{-1}$ (shown as the shaded region).}
    \label{fig:conjecture_plot}
\end{figure}

\section{Estimating $\fnl$}
\label{sec:mcmc}

In this section, we apply our neural network to simulations with $\fnl\ne 0$.
We use a test set of 10 $N$-body simulations with $\sigma_8=0.834$, $\fnl=250$, and non-Gaussian initial conditions generated using the Zeldovich approximation.

\vskip5pt
\noindent
{\bf Power spectra:}
We next verify the predictions in Eqs.\ (\ref{eq:P_m_pi})--(\ref{eq:bngpi_general}) for the large-scale power spectra $P_{m\pi}$, $P_{\pi\pi}$ in an $\fnl$ cosmology.
First, we note that since $\bar\pi = \sigma_8$ (Fig.\ \ref{fig:sigma8_histograms}), our prediction (\ref{eq:bngpi_general}) for the non-Gaussian bias $b^{NG}_\pi$ is:
\begin{equation}
b^{NG}_\pi = 2 \frac{\partial\bar\pi}{\partial\log\sigma_8} = 2 \sigma_8 \label{eq:bngpi_specific}
\end{equation}
and so our prediction (\ref{eq:bpi_general}) for the total bias $b_\pi(k)$ is:
\begin{equation}
b_\pi(k) = b_\pi^G + 2 \sigma_8 \frac{\fnl}{\alpha(k,z)}  \label{eq:bpi_specific}
\end{equation}
In the the top panel of Fig.\ \ref{fig:conjecture_plot}, we compare the bias model (\ref{eq:bpi_specific}) to the empirical bias obtained from cross-correlating $\pi$ and $\delta_m$ in $k$-bins, and find good agreement.
In the bottom panel of Fig. \ref{fig:conjecture_plot}, we verify the prediction that the noise power spectrum $N_\pi$ defined in Eq.\ (\ref{eq:P_pi_pi}) is constant in $k$, by plotting the power spectrum of the residual field $\epsilon(\k) = \pi(\k) - b_\pi(k) \delta_m(\k)$.

We emphasize that simulations with $\fnl\ne 0$ were never seen during the training process.
The predictions in Eqs.\ (\ref{eq:P_m_pi})--(\ref{eq:bngpi_general}) for power spectra in an $\fnl$ cosmology (in particular the prediction $b_\pi^{NG} = 2\sigma_8$) are based entirely on the NN response to varying $\sigma_8$, and general considerations of locality.
Therefore, the verification of these predictions is a strong test of our formalism.

\vskip5pt
\noindent
{\bf MCMC pipeline:}
Now that our model for the power spectra $P_{m\pi}$, $P_{\pi\pi}$ has been verified, we develop an MCMC pipeline that combines large-scale modes of $\pi(\k)$ and $\delta_m(\k)$ in a joint analysis.

\begin{figure*}
     \begin{subfigure}%
         \centering
         \includegraphics[width=0.49\linewidth]{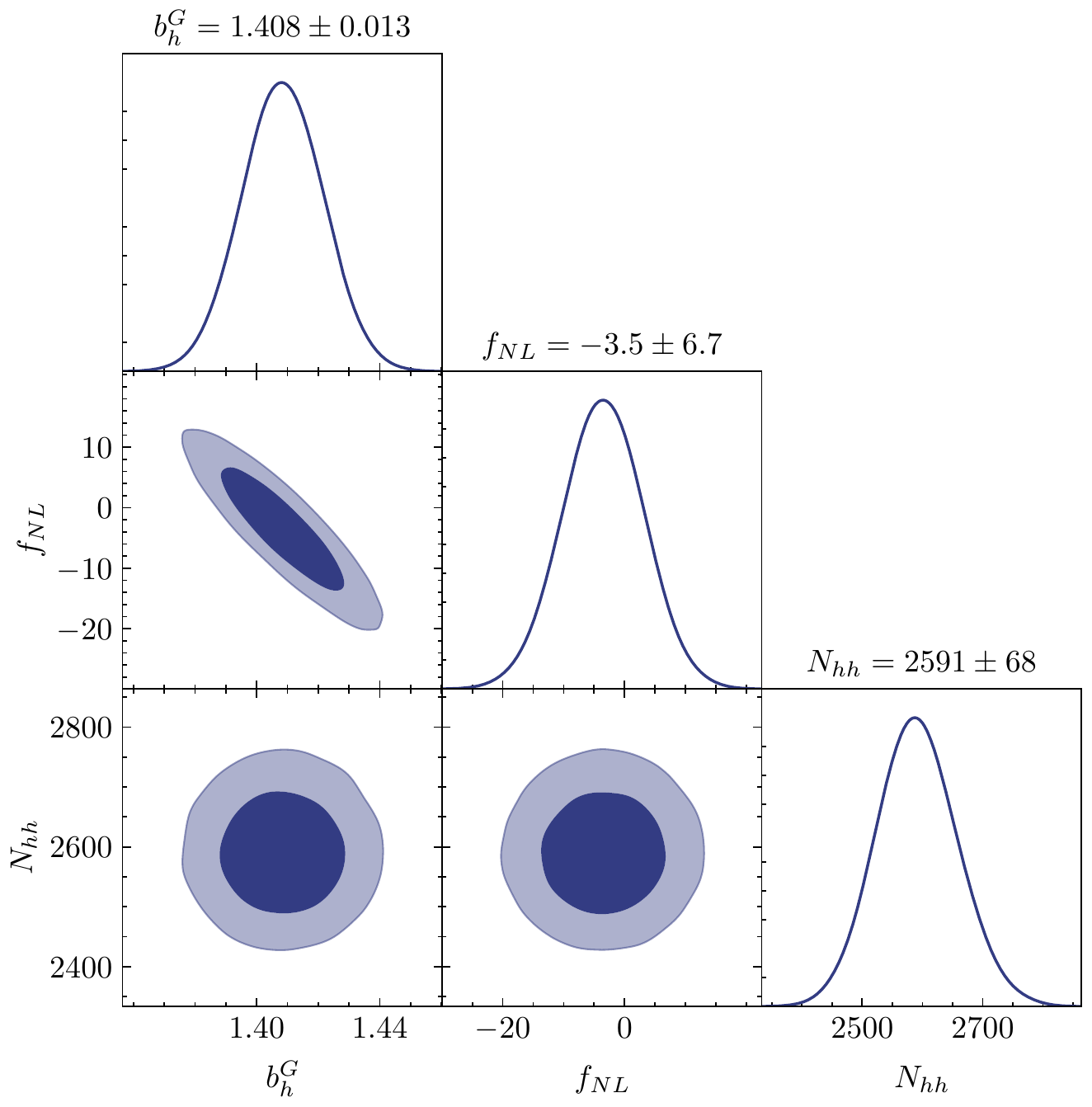}
         \label{fnl_constraint_halo}
     \end{subfigure}
     \begin{subfigure}%
         \centering
         \includegraphics[width=0.49\linewidth]{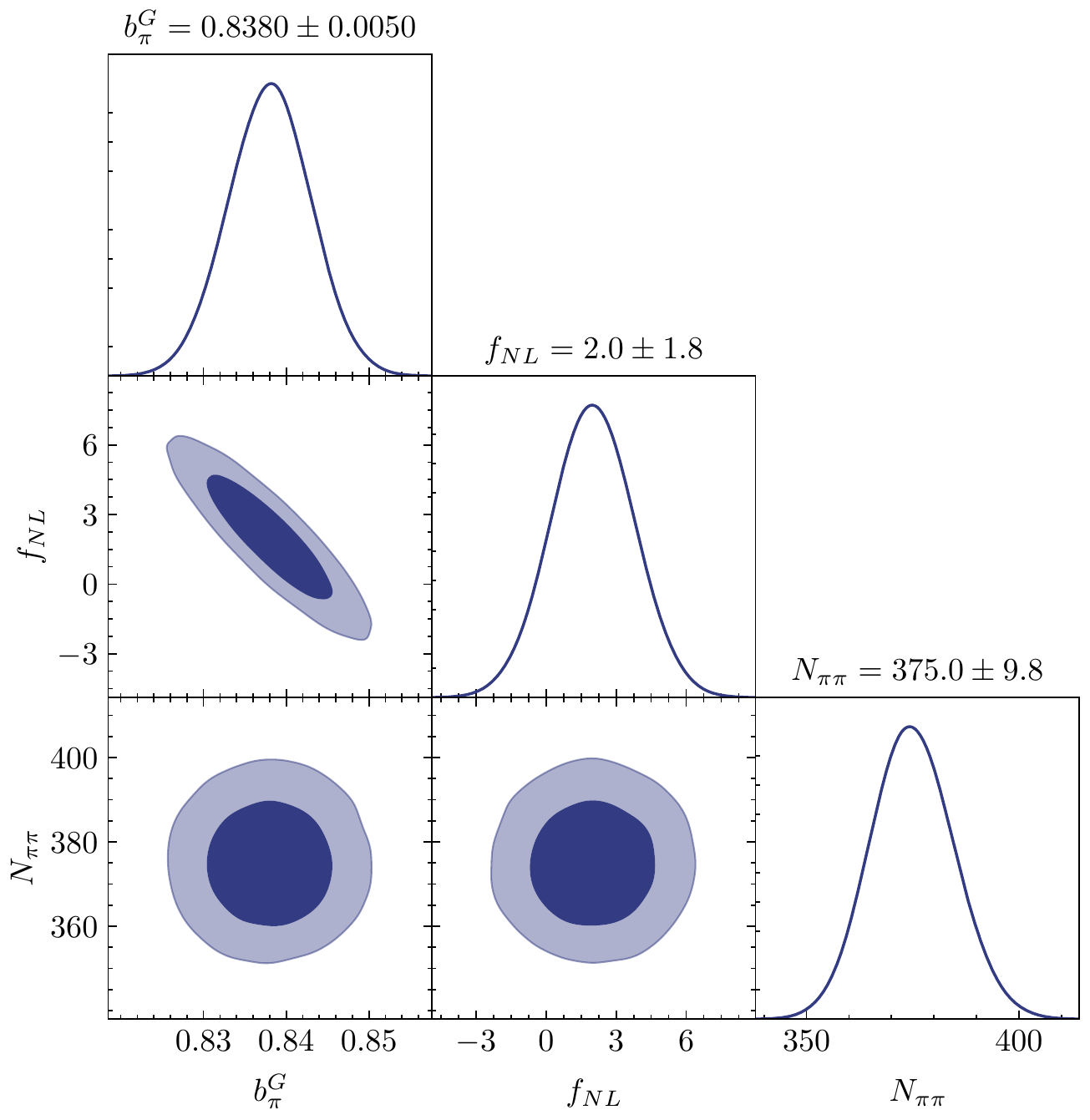}
         \label{fig:fnl_constraint_pi}
     \end{subfigure}
     \caption{MCMC posteriors on $\fnl$ and nuisance parameters (either $(b^G_h,N_{hh})$ or $(b^G_\pi,N_{\pi\pi})$) from joint analysis of 100 Quijote simulations with $\fnl = 0$. 
      \textit{Left.} Traditional halo based analysis using large-scale modes of the matter field $\delta_m(\k)$ and halo field $\delta_h(\k)$.
      \textit{Right.} Neural network based analysis using $\delta_m(\k)$ and the NN output field $\pi(\k)$. The neural network reduces the error bar on $\fnl$ by a factor $\sim$3.5.}
     \label{fig:fnl_constraint}
\end{figure*}

\begin{figure}
    \centering
    \includegraphics[width=0.48\textwidth]{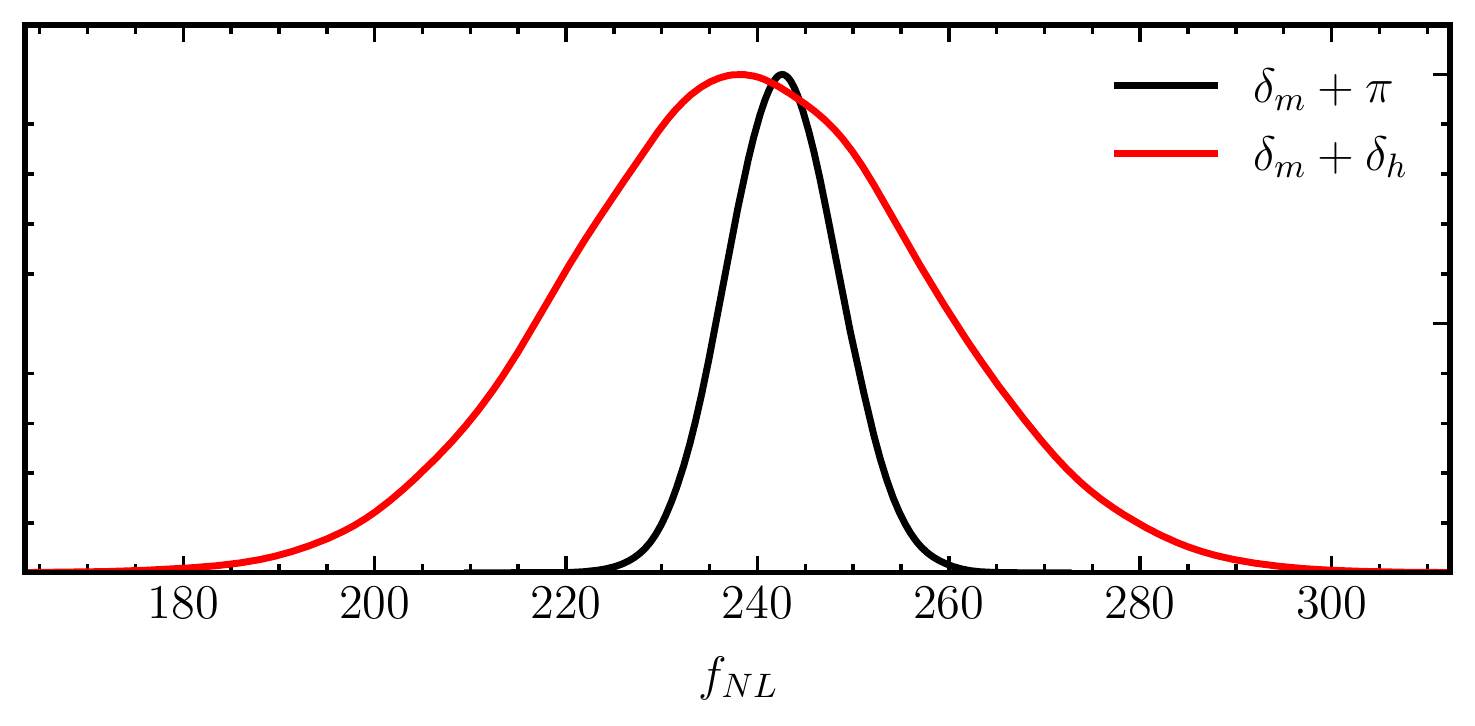}
    \caption{MCMC posterior on $\fnl$ from joint analysis of 10 $N$-body simulations with $\fnl=250$, using either the matter+halo fields (red), or matter+$\pi$ fields (black), where $\pi(\x)$ is the NN output field.
    The 1-d $\fnl$ likelihoods are marginalized over nuisance parameters (either $(b^G_h,N_{hh})$ or $(b^G_\pi,N_{\pi\pi})$).}
    \label{fig:fnl250_posterior}
\end{figure}

\comment{
\vskip5pt
\noindent
{\bf Model:} For $N$-body simulations with Gaussian initial condition, the field $\pi_i$ on large scales can be expressed using leading-order perturbative bias \footnote{Higher-order and non-local contributions are negligible for case in hand.} expansion
\begin{equation}
    \pi_i = b_\pi \delta + \epsilon_i 
\end{equation}
where $\epsilon_i$ is random gaussian noise uncorrelated with $\delta$ with power spectrum defined by
\begin{equation}
    \langle \epsilon_i(k) \epsilon_i(k`) \rangle = (2\pi)^{3} \delta^3(k-k`)N_{\pi_i\pi_i}
\end{equation}
The bias expansion has a simple physical interpretation under the \textit{peak-background split} formalism where the bias is
\begin{equation}
    b_{\pi_i} = \frac{\partial \bar{\pi}_i}{\partial \delta_l}
\end{equation}
where $\delta_l$ is large-wavelength matter mode such that $\delta = \delta_l + \delta_s$. \\
\noindent In an $f_{nl}\neq 0$ universe, the expansion becomes
\begin{equation}
    \pi_i = \Bigg(b_{\pi_i} + \frac{2\beta_{\pi_i} f_{nl}}{\alpha}\Bigg)\delta_m + \epsilon_i 
\end{equation}
where $\beta_{{\pi}_i}$ is the non-Gaussian bias of the $\pi$ field given under PBS by 
\begin{equation}
    \beta_{\pi_i} = \frac{\partial \ln \bar{\pi}_i }{\partial \ln \sigma_8}
\end{equation}
\utkarsh{add remaining details in appendix??}
}

The Gaussian likelihood for our data vector $\mathcal{D}=[\delta_m$, $\pi]$ given model parameters $\Theta = (\fnl, b^G_\pi, N_{\pi\pi})$ is:

\begin{equation}
    \mathcal{L}(\Theta|\mathcal{D}) \propto \prod_{k}
    \frac{1}{\sqrt{\mbox{Det}\, C(k)}}
    \exp(-\frac{{\mathcal D}(\k)^{\dagger} C(k)^{-1} {\mathcal D}(\k)}{2V}) \nonumber
\end{equation}
where the $2\times2$ covariance matrix $C(k)$ is:
\begin{align}
C(k) &= 
\begin{bmatrix}
P_{mm}(k) & P_{m\pi}(k)\\
    P_{m\pi}(k) &  P_{\pi \pi}(k) %
\end{bmatrix} \nonumber \\
&=
\begin{bmatrix}
P_{mm}(k) & b_\pi(k) P_{mm}(k)\\
    b_\pi(k) P_{mm}(k) & b_\pi(k)^2 P_{mm}(k) + N_{\pi\pi} 
\end{bmatrix}
\end{align}
with $b_\pi(k)$ given by Eq.\ (\ref{eq:bpi_specific}).
We truncate the likelihood at $k_{\rm max}=0.014$ $h\, \mathrm{Mpc^{-1}}$. The posterior is defined using flat priors over a reasonable range for the model parameters $\Theta$. We sample the posterior using affine-invariant sampling implemented in \texttt{emcee}\cite{Foreman_Mackey_2013} to obtain constraints on $\fnl$.

To compare the neural network to a traditional halo based analysis, we also run our MCMC pipeline using the halo field $\delta_h(\k)$ instead of the NN-derived field $\pi(\k)$. The only change is that we replace $b^{NG}_\pi = 2 \sigma_8$ by the non-Gaussian halo bias $b^{NG}_h$, which we measure in simulations using Eq.\ (\ref{eq:bh_ng}).

\vskip5pt
\noindent \textbf{MCMC results:} 
We begin by analysing $N$-body simulations with Gaussian initial conditions \ie\ $\fnl=0$. %
We jointly analyze 100 \texttt{fiducial} Quijote simulations by multiplying together their posteriors before sampling. 
In the right panel of Fig.\ \ref{fig:fnl_constraint}, we show $\fnl$ constraints from a joint analysis of the large-scale matter density $\delta_m(\k)$ and the NN-derived field $\pi(\x)$.
In the left panel, we show a similar analysis using $\delta_m(\k)$ and the {\em halo} field $\delta_h(\k)$.
In both cases, the result is consistent with $\fnl=0$ as expected.
However, the neural network gives an $\fnl$ error which is 3.5 times better than the halo based analysis!

In the left panel of Figure \ref{fig:fnl_constraint}, we used a single halo field consisting of all halos with $\ge 20$ particles ($M_{\rm min} = 1.3 \times 10^{13}$ $h^{-1}\,M_\odot$).
We checked that if narrow halo mass bins are used with optimal weighting, the Fisher forecasted error $\sigma(\fnl)$ is only 25\% better than the single-bin case.

In Fig.\ \ref{fig:fnl250_posterior}, we show $\fnl$ constraints from a joint MCMC analysis of 10 simulations with $\fnl=250$.
We can see that for these non-Gaussian simulations, the correct value of $\fnl$ is recovered, and the NN improvement over halos is just as good as in the $\fnl=0$ case.   %

\vskip5pt
\noindent \textbf{Robustness:} 
To frame the issue of robustness concretely, imagine that the small-scale astrophysics in the real universe is slightly different from the training set.
How will our constraints on $\fnl$ be affected?

Since the parameters $(b^G_\pi, b_\pi^{NG}, N_\pi)$ are sensitive to small-scale physics, their values will differ slightly from the training set.
For $b^G_\pi$ and $N_\pi$, this is harmless since we marginalize these parameters in our MCMC anyway.

For $b^{NG}_\pi$, we note that in the formalism from \S\ref{sec:formalism}, the parameters $b^{NG}_\pi$ and $\fnl$ only appear in the combination $(b^{NG}_\pi \fnl)$.
Therefore, a small change in $b^{NG}_\pi$ is equivalent to a change in the normalization of $\fnl$ -- it cannot ``fake'' a detection of nonzero $\fnl$.
Physically, this is because local physics cannot generate a term in the bias $b_\pi(k)$ proportional to $1/k^2$.
This is qualitatively similar to the familar case of halo counts.

Our method does depend on having a rough estimate for $b^{NG}_\pi$ based on training data.
As a check, we estimated the cosmological parameter dependence of $b^{NG}_\pi$ using the Quijote ``latin hypercube'' simulations, which vary cosmological parameters over wide ranges.
We find that $\bar\pi$ is well modelled by a quadratic polynomial in $(\Omega_m,\Omega_b,h,n_s,\sigma_8)$.
Using this quadratic model, we find that if cosmological parameters are varied within Planck+BAO $2\sigma$ errors \cite{Planck:2018vyg}, the change in $b^{NG}_\pi = 2 (\partial\bar\pi/\partial\log\sigma_8)$ is $\le 1$\%.
In future work, we hope to extend this analysis to study dependence of $b^{NG}_\pi$ on subgrid physics.

\section{Conclusion and Outlook} 
\label{sec:conclusion}

In this letter we have demonstrated that the statistical power of neural networks can be combined with
the idea of $(1/k^2)$ non-Gaussian bias
to arrive at a robust measurement of $\fnl$ from the matter distribution.
Unlike forward modelling approaches which are difficult at strongly non-linear scales, our approach can use information from very small scales and still remain robust. 

Our main next step will be to quantify to what extent the method presented here can improve $\fnl$ constraints from realistic galaxy surveys, 
rather than the matter field. 
Machine learning based $\sigma_8$ constraints from simulated galaxy distributions have been examined in \cite{Ntampaka_2020} (using a halo occupation distribution), \cite{Villanueva-Domingo:2022rvn} (using the hydrodynamic CAMELS simulations) and \cite{Perez:2022nlv} (using a semi-analytic galaxy formation model). In particular, \cite{Villanueva-Domingo:2022rvn} highlighted the problem that different baryonic subgrid models lead to inconsistent results, which is precisely the issue our method is designed to overcome for $f_{NL}$. Recently, \cite{Valogiannis:2021chp} introduced a non-linear estimator based on the Wavelet Scattering Transform (WST) and even applied it to BOSS data \cite{Valogiannis:2022xwu} to extract cosmological parameters including $\sigma_8$. The WST behaves similarly to a neural network and thus the claimed improvements in $\sigma_8$ suggest that our $\fnl$ method could also work well for galaxies. We will investigate this question in detail in upcoming work.

A straightforward generalization of our method is to other scale-dependent biases, such as those induced by the trispectrum $g_{NL}$ parameter  \cite{Smith_2012}, neutrino masses \cite{Chiang:2017vuk}, and isocurvature perturbations \cite{Barreira:2019qdl}.
More generally, our approach of using a neural network as a local probe 
may generalize to other observables which are large-scale modulations of local non-Gaussian fields or cross-correlations. This is a common setup in cosmology, often exploited for quadratic estimators.   \\

\section*{Acknowledgements} 

Part of this work was performed at the Aspen Center for Physics, which is supported by National Science Foundation grant PHY-1607611.
MM acknowledges support from DOE grant DE-SC0022342. 
KMS was supported by an NSERC Discovery Grant and a CIFAR fellowship.
Research at Perimeter Institute is supported in part by the Government of Canada through the Department of Innovation, Science and Economic Development Canada and by the Province of Ontario through the Ministry of Colleges and Universities.
Perimeter Institutes’s HPC system ``Symmetry'' was used to perform some of the analysis presented in the letter. We have extensively used several python libraries including \texttt{numpy}\cite{harris2020array}, \texttt{matplotlib}\cite{Hunter:2007}, \texttt{CLASS}\cite{Blas_2011}, \texttt{getdist}\cite{Lewis:2019xzd} and \texttt{SciencePlots}\cite{SciencePlots}.
\bibliography{sigmafnl}

\end{document}